%% file: BitcoinBackbone.tex
\pdfoutput=1 
\documentclass[conference,compsoc]{IEEEtran}

\usepackage[bookmarks=false]{hyperref}
\usepackage[left=1.2in,top=1.2in,right=1.2in,bottom=1.5in, portrait]{geometry}
\usepackage[english]{babel}
\usepackage[utf8]{inputenc}
\usepackage[export]{adjustbox} 
\usepackage{pgfplots}
\pgfplotsset{width=10cm,compat=1.9}
\usepackage{dirtytalk}
\usepackage{amsmath}
\usepackage{amssymb}
\usepackage{hyperref}
\usepackage{multirow}
\usepackage{listings}
\usepackage[noend]{algpseudocode}
\usepackage{float}
\usepackage{graphicx}
\usepackage[colorinlistoftodos]{todonotes}
\usepackage{subcaption}
\usepackage[utf8]{inputenc}
\usepackage[english]{babel}
\usepackage{geometry}
\usepackage{comment}
\usepackage{microtype}

\usepackage[ruled,vlined]{algorithm2e}
\include{mymacro}

\usepackage[backend=biber,
style=alphabetic,
sorting=ynt]{biblatex}
\usepackage{enumitem}
\begin{document}

\title{Model Checking Bitcoin and other Proof-of-Work Consensus Protocols}

\author{\IEEEauthorblockN
{Max DiGiacomo-Castillo\IEEEauthorrefmark{1},
Yiyun Liang\IEEEauthorrefmark{2},
Advay Pal \IEEEauthorrefmark{3}, 
John C. Mitchell \IEEEauthorrefmark{4}}

\IEEEauthorblockA{Department of Computer Science,
Stanford University\\
Email: \IEEEauthorrefmark{1}maxdigi@stanford.edu,
\IEEEauthorrefmark{2}isaliang@stanford.edu,
\IEEEauthorrefmark{3}advaypal@stanford.edu,
\IEEEauthorrefmark{4}jcm@stanford.edu}}

\maketitle

\begin{abstract}
The Bitcoin Backbone Protocol \cite{garay_bitcoin_2015} is an abstraction of the bitcoin proof-of-work consensus protocol. 
We use a model-checking tool (UPPAAL-SMC) to examine the concrete security of proof-of-work consensus by
varying protocol parameters and using an adversary that leverages the selfish mining strategy introduced in \cite{garay_bitcoin_2015}. 
We provide insights into modeling proof-of-work protocols and demonstrate tradeoffs between operating parameters. 
Applying this methodology to protocol design options, we show that the uniform tie-breaking rule 
from \cite{eyal_majority_2018} decreases the failure rate of the chain quality property, 
but increases the failure rate of the common prefix property. 
This tradeoff illustrates how design decisions affect protocol properties, within a range of concrete operating conditions, 
in a manner that is not evident from prior asymptotic analysis.
\end{abstract}

\IEEEpeerreviewmaketitle

\section{Introduction}
Bitcoin is widely used, with a current market capitalization of over 120 billion USD.
Given the massive level of economic activity and the potential for future growth, it is natural to ask: 
how confident can we be in the security and reliability of this system and its variants?
While practical experience suggests that Bitcoin and related systems are robust against various kinds of misuse,
their security rest on a combination of features whose interactions are not fully understood.
For example, could advances in computing change the balance of power between honest miners and  adversaries?

Initial studies of blockchain security \cite{garay_bitcoin_2015, li_continuous-time_2020} 
have proved some fundamental relationships involving security parameters and the probabilities of desired properties.
In this paper, we develop a formal model based on the \emph{Bitcoin Backbone Protocol} abstraction \cite{garay_bitcoin_2015}
and use a statistical model checking tool (UPPAAL-SMC) to study its security. 
We focus on how the properties of the backbone protocol vary as a function of concrete parameters,
in a network where an adversary is capable of selfish mining (delaying the release of malicious blocks). 

The main contributions of this paper are:
\begin{itemize}[leftmargin=0.25cm]
    \item We demonstrate a way to model the backbone protocol in the presence of a selfish-mining adversary.
    \item We quantitatively analyze a concrete trade-off between different security properties, based on how honest miners act on receiving new blocks.
    \item We demonstrate how the failure rate of the backbone properties vary with different values of $f$, where $f$ is the probability that at least one honest party mines a block in a round, a parameter that is different for different cryptocurrencies.
\end{itemize}

The paper is divided into the following sections. 
In Section 2, we provide an overview of the Bitcoin Backbone Protocol, which is the basis for our UPPAAL-SMC model. 
In Section 3 we describe how we model the backbone protocol using the tool. 
In Section 4 we describe our results.
Finally, Section 5 provides an overview of related work, Section 6  outlines potential future work, and Section 7 concludes.

\section{Overview of the Backbone Protocol}

The Bitcoin Backbone Protocol was introduced in \cite{garay_bitcoin_2015} and subsequently improved in 
\cite{journals/iacr/GarayKL16, li_continuous-time_2020}. 
The aim of the backbone protocol is to analyze the core mechanisms of proof-of-work consensus protocols 
and provide more detailed security guarantees than those provided by Nakamoto's whitepaper \cite{nakamoto2008bitcoin}. 
The protocol captures key elements of Bitcoin and related Proof-of-Work (PoW) consensus protocols that are use in other blockchains. 

The protocol represents time as a series of discrete \textit{rounds,} each short enough so that
the probability of any party completing the work needed to write a new block is low.
Each round proceeds as follows:
\begin{enumerate}
    \item \textit{Start:} Each miner starts the round with a preferred current chain.
    \item \textit{Check:} Each miner begins the round by checking for new chains.
    \item \textit{Adopt:} Each honest miner adopts the best chain visible to it, using a selection criteria 
    that is defined by the blockchain. 
    \item \textit{Mine:} Each miner queries a cryptographic hash function. If they probabilistically succeed with proof of work in this round, they append a block to their chain. 
    \item \textit{Broadcast:} Any miner that modifies its local chain  will broadcast its new chain to other parties. 
\end{enumerate}

In modeling the backbone protocol, a party may be designated as honest or adversarial. 
Honest parties will immediately share blocks they find, and select their chain based on the protocol’s designated chain selection algorithm. They will not deviate from the steps outlined above.

The protocol adversary represents a possible coalition of malicious miners.
The adversary is therefore able to query the cryptographic hash function and produce blocks, 
with a success probability per round that may differ from a single honest miner. 
In addition, a possible network advantage of the adversary is represented by allowing the adversary 
to inject messages into any miner's input channel and reorder any input channel 
at will.\footnote{We use "input channel" to mean the same thing as "RECEIVE()" from \cite{garay_bitcoin_2015}.} 
The adversary can also select a preferred chain arbitrarily (rather than using the honest selection criteria given in step 3 above)
and withhold blocks in order to transmit them at a later round.

In \cite{garay_bitcoin_2015}, the authors make certain assumptions about the system:
\begin{itemize}[leftmargin=0.25cm]
    \item The protocol is executed by a fixed number of parties \textit{n}. 
    \item Parties do not know the source of messages.
    \item 
    All messages are delivered by the end of a round.\footnote{This assumption is for the synchronous model of \cite{garay_bitcoin_2015}.}
    \item All parties involved are allowed the same number of queries to a cryptographic hash function, for the PoW computation. 

\end{itemize}

The protocol parameters from \cite{garay_bitcoin_2015} that are relevant to our work are
shown in Figure \ref{fig:parameters}; the reader may consult \cite{garay_bitcoin_2015} for
further detail.
\begin{figure}[h!]
    \centering
    \fbox{\begin{minipage}{{0.5\textwidth}}
    \scriptsize
$\lambda$: security parameter\\
\textit{n}: number of parties \\
\textit{t}: number of parties controlled by the adversary\\
\textit{f}: probability at least one honest party finds a PoW in a round\\
$\epsilon$: concentration of random variables in typical executions\\
$\mu$: proportion of blocks in honest chains that were mined by honest parties\\
\textit{k}: number of blocks for common prefix\\
$\ell$: number of blocks for chain quality \\
\textit{s}: rounds for chain growth property \\ 
$\tau:$ chain growth parameter 
\end{minipage}}
    \caption{Positive integers \textit{n}, \textit{t}, \textit{s}, $\ell$, \textit{k}; positive reals \textit{f}, $\epsilon, \mu, \tau, \lambda$, with $\textit{f}, \epsilon, \mu \in (0,1).$ }
    \label{fig:parameters}
\end{figure}

If the modeling assumptions listed above hold and the majority of parties are honest, the behavior of the protocol can be described using 
the concept of typical execution, introduced in \cite{garay_bitcoin_2015}.
By definition, a \textit{typical execution} is a sequence of rounds in which the random variables are close to their expected values. 
A straightforward calculation shows that a typical execution occurs with probability $1-e^{-\Omega({\epsilon}^2\lambda\textit{f})}$. 

The desired properties of the backbone protocol are:

\begin{itemize}[leftmargin=0.25cm]
    \item \textit{Chain Quality:} Chain quality is the proportion of honest blocks in the chain of an honest participant. A subsection of $\ell$ blocks in a honest chain will have at least $\mu\ell$ blocks that were mined by honest parties. 
    Note that $\ell \geq 2\lambda\textit{f}$ for provable security. 
    
      

    \item \textit{Common Prefix:} The common prefix property holds if honest parties that prune $k$ blocks from their chain share a common view with another honest party. 
    In a typical execution the common prefix holds for \textit{k} $\geq$ 2$\lambda$\textit{f}. 
    
    
    \item \textit{Chain Growth:} The chain growth property measures  how quickly the chains of honest parties grow. Given that $\tau$ = (1 - $\epsilon$)\textit{f}, honest chains grow at least as fast as $\tau$ $\cdot$ \textit{s} in a typical execution. Note that \textit{s} $\geq \lambda$. 
\end{itemize}

The authors of \cite{garay_bitcoin_2015} show that if these properties hold, then persistence and liveness, which are crucial to the robustness of the system, also hold. Persistence states that once a transaction reported by an honest party becomes ‘deep enough’ in a blockchain, it is present in the chain of every other honest party at the same position. Liveness states that any transaction that comes from an honest party, and is provided to all other honest parties, will be inserted into all honest ledgers. The authors note that persistence and liveness are not proof that Bitcoin meets all of its objectives, as the analysis assumes the number of parties is fixed and there is an honest majority. 

\cite{garay_bitcoin_2015} also provides a similar analysis for a network setting which is not highly synchronous, meaning there is an upper bound on the amount of rounds a message takes to be delivered. We omit a discussion of this because it is beyond the scope of our study, but extending our analysis to this model remains a potential direction for future work.

\section{Modeling the Backbone Protocol}
In this section we introduce our model checking formalism of the backbone protocol. We first define the following for each participant in the protocol:

\begin{itemize}[leftmargin=0.25cm]
    \item Input channel: Chains from other participants will be sent to a participant's input channel.
    \item Output channel: When a participant successfully mines a block, it sends its newly mined block to its output channel, to be broadcasted to the rest of the network.
\end{itemize}

We now present how honest and adversarial participants are modeled in the protocol. Honest participants are modeled by Algorithm \ref{algo:honest_mining} (above), and the adversary is modeled by Algorithm \ref{algo:adv_mining} (above).

\begin{algorithm}[t]
\scriptsize
\KwIn{Honest miner m}
	\caption{Honest Mining} 
	    check input channel and update bestBlock[m]
		
		\If{Mining success} {
		  create newBlock
          
          publish newBlock with newBlock $\mapsto$ bestBlock[m]
          
          set bestBlock[m] = newBlock
        }
\caption{Honest Mining}
\label{algo:honest_mining}
\end{algorithm}

\begin{algorithm}[t]
\scriptsize
\KwIn{Adversarial miner a}
		decide strategy and update arbitraryBlock[a]
        
        \If{advMiningSuccess} {
          create badBlock $\mapsto$ arbitraryBlock[a]
          
          publish badBlock or keep private
        }
        
        route published and private blocks to honest miners, at will
\caption{Adversarial Mining}
\label{algo:adv_mining}
\end{algorithm}

As shown in Algorithm \ref{algo:honest_mining}, $bestBlock$ is a global array that stores the head of each miner's blockchain. At the beginning of each round, an honest miner checks its input channel, checking its local chain against any chains published in the network. The miner then attempts to mine a block. If mining was successful, a new block will be created and published to the rest of the network. The miner will also update its local blockchain by appending its newly-mined block.

In contrast to Algorithm \ref{algo:honest_mining}, Algorithm \ref{algo:adv_mining} allows an adversarial miner to decide the best strategy to use in the current round, and keep blocks private \footnote{Private blocks are not broadcasted to other miners.}.

In the first execution round, everyone adopts the genesis block. In all subsequent rounds, honest participants select their block by referring to their input channels. The adversary may select their block arbitrarily. 

\begin{figure}[t]
    \centering
  \includegraphics[width=0.5\textwidth, height = 5cm]{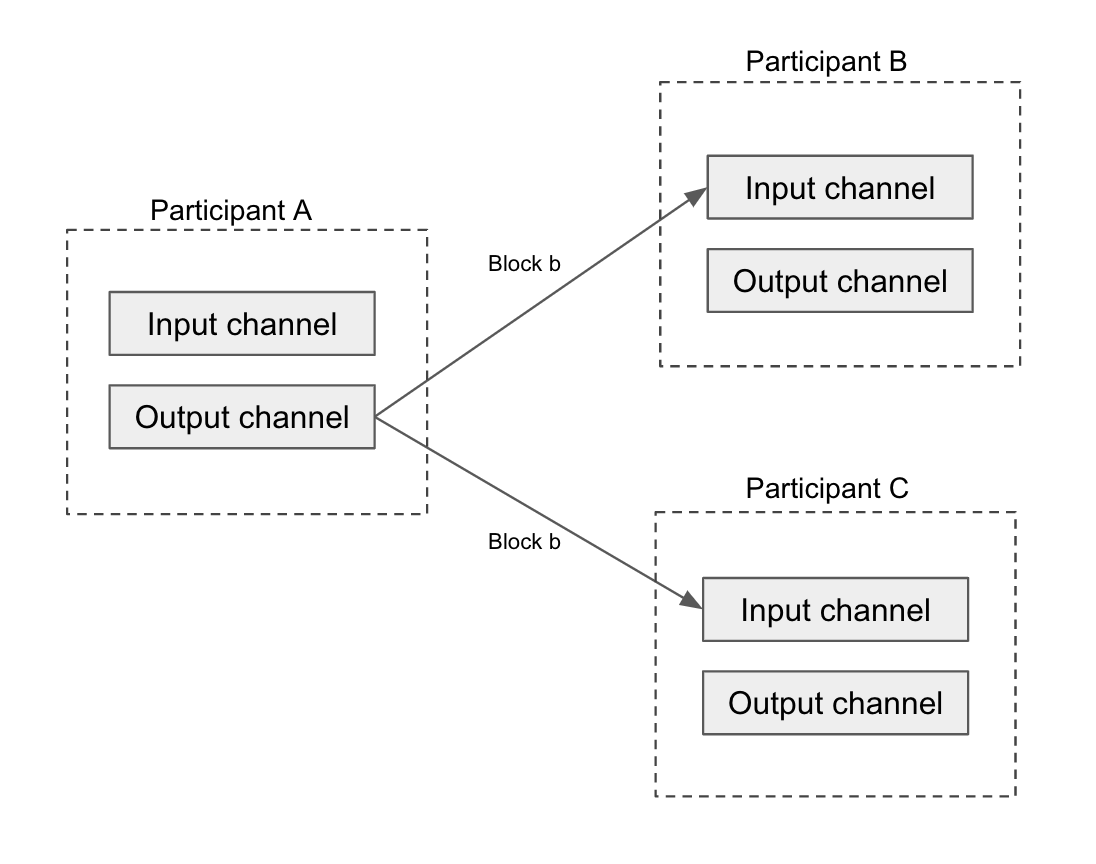}
  \caption{Participant \textit{A}  broadcasts a block}
  \label{fig:broadcast}
\end{figure}

Figure \ref{fig:broadcast} shows how the input and output channels are updated at each round. Assume \textit{A}, \textit{B}, and \textit{C} are honest miners and \textit{A} successfully mined a block in the most recent round. Participant \textit{A} will send its block to its output channel.  The block is then sent to the input channel of all other participants on the network. In the following round, the honest parties will check their input channels and use a selection algorithm to determine whether to adopt \textit{A's} block.

In the presence of an adversary, input channels can be manipulated. The adversary can standby  until all honest parties have completed mining. This allows the adversary to maximize the amount of information they can account for when deciding their propagation strategy.  

\begin{figure}[t]
\centering
  \includegraphics[width=0.5\textwidth]{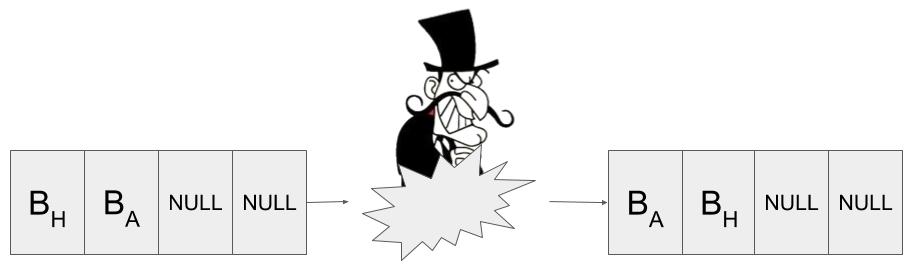}
  \caption{Example of an adversary reorganizing an input channel. $B_H$ is a block mined by an honest party and $B_A$ was mined by the adversary.}
  \label{fig:adversary}
\end{figure}

When all honest miners have completed a round, the adversary checks each participant's output channel for blocks and set the order of blocks on each honest input channel (Figure \ref{fig:adversary}). By default, honest parties will adopt the first chain they receive when encountering chains of equal length. This allows the adversary to rearrange input channels to win head-to-head ties. The adversary also decides whether or not to share blocks they have mined with honest participants. 

At the end of this process, the output channels of each participant will include any blocks not sent to the input channels by the adversary, but mined by honest participants in the most recent round. At this point all parties have attempted to solve the PoW, and been informed of all blocks found in the most recent round. The participants continue the process above for a finite-amount of rounds.  

\subsection{Differences from Backbone Protocol}

We now review the key ways that our UPPAAL-SMC model differs from the backbone protocol. 

\begin{itemize}[leftmargin=0.25cm]
\item Parties: We model the adversary as one party. The adversary has a mining power of $\alpha$ and the honest parties collectively have a mining power of (1-$\alpha$). This is equivalent to \cite{garay_bitcoin_2015}, where the adversary controls a subset of honest parties.
\item Transactions: We omit the inclusion of transactions in blocks. This does not affect the outcome of the backbone properties. 
\item Mining: Our mining process relies on  probabilistic transitions to either a success or failure state. This avoids modeling hash functions. Our work does not account for honest mining pools or a single party mining multiple blocks in one round.
 \item Message Propagation: We assume that all messages will be delivered at the \textit{end} of a round. This means no blocks are sent or received during the mining phase of a round. 
 \item Non-Determinism: Whereas the Bitcoin Backbone Protocol quantifies over all possible adversaries, we only model a selfish mining adversary. This is because UPPAAL-SMC does not handle non-determinism. 
\end{itemize}

\section{Results and Analysis}

We now introduce the results we obtained. UPPAAL-SMC simulates many runs of our system, and then computes the probability of a property holding. We use a 95\% confidence interval. This means that 95\% of the time the true probability of a failure event is within $\pm0.05$ of the value computed. 
 
\subsection{Fork Resolution Rules}
In Bitcoin, when honest miners receive multiple longest chains, they adopt the one they received first. We refer to this fork resolution rule as the \say{First-Received Rule}. Under the first-received rule, a well connected adversary can gain a large advantage. If they are able to route their blocks quickly, more parties will adopt their chain. The assumption in \cite{garay_bitcoin_2015} is that all honest participants, except for one, will adopt the adversary's preferred block in the case of a fork. We also make this assumption.

We begin by testing uniform tie breaking, an alternative where an honest party adopts one longest chain at random.  
This modification can limit the network advantage that an adversary can obtain. No matter the adversary's propagation strategy, uniform tie breaking will cause about 50\% of parties to adopt the block preferred by the adversary.

For each resolution rule, we use our model to measure the probability of each backbone property failing. For the following experiments, we let $n = 8, \alpha = 0.33 , f = 0.02, \text{and } \mu =0.39$.  Like \cite{garay_bitcoin_2015}, we assume $f \approx 0.02$   for Bitcoin.


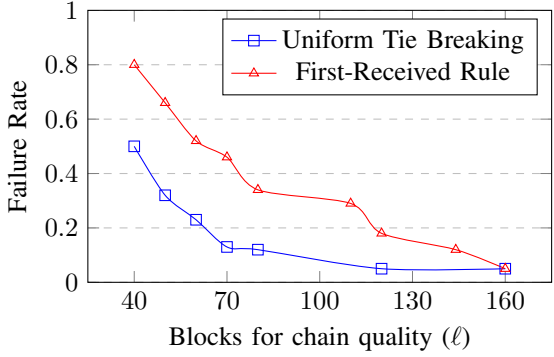
\begin{figure}[t]
\centering
\begin{tikzpicture}
\begin{axis}[
    title={},
    xlabel={Blocks for chain quality ($\ell$)},
    ylabel={Failure Rate},
    xmin=25, xmax=175,
    ymin=0, ymax=1,
    xtick={40,70,100,130,160},
    ytick={0,.20,.40,.60,.80,1},
    legend pos= north east,
    ymajorgrids=true,
    grid style=dashed,
    width=0.5\textwidth,
    height = 5.2cm,
]
\addplot+[smooth][
    color=blue,
    mark=square,
    ]
    coordinates {
    (40, 0.50)
    (50, 0.32)
    (60, 0.23)
    (70, 0.13)
    (80, 0.12)
    (120, 0.05)
    (160, 0.05)
    };
\addplot+[smooth][
    color=red,
    mark=triangle,
    ]
    coordinates {
    (40, 0.80) 
    (50, 0.66)
    (60, 0.52)
    (70, 0.46)
    (80, 0.34)
    (110, 0.29)
    (120, 0.18)
    (144, 0.12)
    (160, 0.05)
    };
    \legend{Uniform Tie Breaking, First-Received Rule }

\end{axis}
\end{tikzpicture}
    \caption{Failure rate of chain quality given two fork resolution rules}
    \label{cq_selection}
\end{figure}

In agreement with \cite{Sapirshtein2015OptimalSM}, we find that the failure rate of the chain quality property is consistently lower when uniform tie breaking is used (Figure \ref{cq_selection}). Still, the failure rate under both resolution rules decreases towards zero for large $\ell$. Large $\ell$ means a longer subsection of blocks is checked against the chain quality property, and prevents the adversary from breaking the property by mining more blocks than expected over a short time period.

\begin{figure}[t]
\centering
\begin{tikzpicture}
\begin{axis}[
    title={},
    xlabel={Blocks for common prefix (\textit{k})},
    ylabel={Failure Rate},
    xmin=3.5, xmax=10.5,
    ymin=0, ymax=1,
    xtick={4,6,8,10,12,14},
    ytick={0,.20,.40,.60,.80,1},
    legend pos=north east,
    ymajorgrids=true,
    grid style=dashed,
    width=0.5\textwidth,
    height = 5.2cm,
]

\addplot+[smooth][
    color=blue,
    mark=square,
    ]
    coordinates {
    (4, 0.95)
    (5, 0.79)
    (6, 0.51)
    (7, 0.29)
    (8, 0.10)
    (10, 0.05)
    };
\addplot+[smooth][
    color=red,
    mark=triangle,
    ]
    coordinates {
    (4, 0.10)
    (5, 0.05)
    (6, 0.05)
    (7, 0.05)
    (8, 0.05)
    (10, 0.05)
    };
\legend{Uniform Tie Breaking, First-Received Rule}
\end{axis}
\end{tikzpicture}
\caption{Failure rate of common prefix given two fork resolution rules}
\label{cp_selection}
\end{figure}
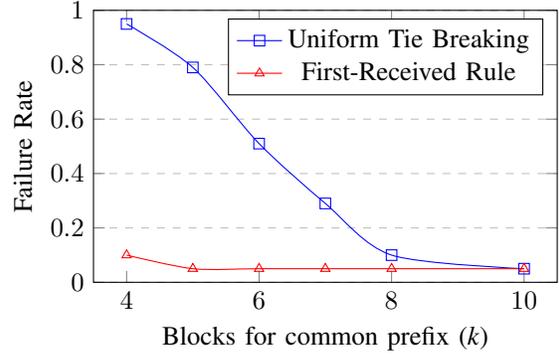

We extend the work of \cite{Sapirshtein2015OptimalSM} and test uniform tie breaking against the common prefix and chain growth properties. Our results show that uniform tie breaking, assuming the adversary follows the selfish mining strategy we implemented, has a negative impact on the common prefix of honest blockchains (Figure \ref{cp_selection}). 

Under uniform tie breaking, the selfish mining strategy from \cite{garay_bitcoin_2015} is similar to the common prefix attack described in \cite{eprint-2015-26827}. Because uniform tie breaking roughly splits the honest participants onto two competing chains, our selfish mining attacker produces long forks with higher probability. Forks are caused when the adversary releases a private block or two honest parties find a block in the same round. The forks take longer to resolve, on average, because the parties equally mine on each branch.

\begin{figure}[t]
\centering
\begin{tikzpicture}
\begin{axis}[
    title={},
    xlabel={Expected blocks ($\tau \cdot$ \textit{s})},
    ylabel={Failure Rate},
    xmin=35, xmax=265,
    ymin=0, ymax=1,
    xtick={50,100,150,200,250},
    ytick={0,.20,.40,.60,.80,1},
    legend pos= north east,
    ymajorgrids=true,
    grid style=dashed,
    width=0.5\textwidth,
    height = 5.2cm,
]
\addplot+[smooth][
    color=blue,
    mark=square,
    ]
    coordinates {
    (19, 0.95)
    (38, 0.95)
    (57,0.89)
    (76,0.79)
    (95,0.70)
    (114,0.60)
    (133, 0.49)
    (152, 0.37)
    (171, 0.31)
    (190, 0.27)
    (228, 0.23)
    (247, 0.07)
    (266, 0.05)
    };
    \addplot+[smooth][
    color=red,
    mark=triangle,
    ]
    coordinates {
    (19, 0.95)
    (38, 0.95)
    (57,0.93)
    (76,0.87)
    (95,0.67)
    (114,0.59)
    (133, 0.48)
    (152, 0.40)
    (171, 0.29)
    (190, 0.28)
    (228, 0.25)
    (247, 0.09)
    (266, 0.05)
    };
    \legend{Uniform Tie Breaking, First-Received Rule}
\end{axis}
\end{tikzpicture}
\caption{Failure rate of chain growth given two fork resolution rules}
\label{fig:cg_selection}
\end{figure}
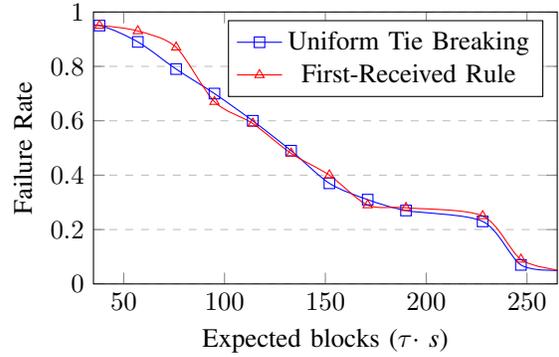

We believe this is a practical implementation of a common prefix attack because it is not clear how an adversary could consistently force honest participants to equally mine on competing chains otherwise. Given that blocks are spread using a gossip protocol, equally dividing the network onto different chains would be difficult.

In \cite{eprint-2015-26827}, the authors state that uniform tie breaking would not help against their common prefix attack. They did not realize the rule could make long-forks attacks easier to execute. Our adversary follows the selfish mining strategy from \cite{garay_bitcoin_2015} and lets uniform tie breaking do the rest. This suggests that protocol modifications to Bitcoin may have unintended consequences when only measured against one property of the backbone.

The chain growth property is also used to compare each resolution rule. We find that over long periods  the difference in performance between the two fork resolution rules with respect to chain growth is negligible (Figure \ref{fig:cg_selection}).

\subsection{Calibrating $f$ for security and speed}

Recall that $f$ is the probability that at least one honest party finds a PoW solution in a round. The importance of calibrating $f$ is made clear by Garay et al \cite{garay_bitcoin_2015}. If the PoW puzzle is too difficult ($f$ is too small) then chain growth suffers. Too few blocks are produced by honest parties, so liveness is hurt. If the PoW puzzle is too easy ($f$ is too large), then the common prefix suffers. There are not enough rounds where only one party finds a PoW solution, so persistence suffers.

Fluctuations in $f$ can be caused by variations in the network's hash rate  or propagation speed. To keep $f$ in a small range, Bitcoin applies a PoW difficulty adjustment every 2016 blocks. The difficulty of PoW is increased or decreased depending on the network’s hash rate. If the hash rate increases, then more blocks are produced within a round ($f$ increases). If the hash rate decreases, then less blocks are produced within a round ($f$ decreases). Bitcoin adjusts its difficulty so that blocks, on average, are produced every 10 minutes. Assuming a full round of propagation takes up to 20 seconds, adjustments keep $f$ between 2-3\% \cite{garay_bitcoin_2015}. Note that this only accounts for changes in the network’s hash rate.


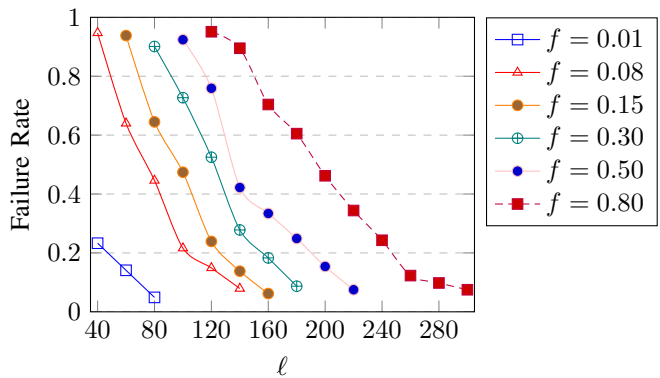
\begin{figure}[h!]
\centering
\begin{tikzpicture}
\begin{axis}[
    title={},
    xlabel={$\ell$},
    ylabel={Failure Rate},
    xmin=35, xmax=305,
    ymin=0, ymax=1,
    xtick={40, 80, 120,  160, 200, 240,  280},
    ytick={0,.20,.40,.60,.80,1 },
    legend pos=outer north east,
    ymajorgrids=true,
    grid style=dashed,
    width=0.9\linewidth,
    height = 5.5cm
]

    \addplot+[smooth][
    color=blue,
    mark=square,
    ]
    coordinates {
    (40, 0.233)
    (60, 0.141)
    (80, 0.049)
    };
    
    \addplot+[smooth][
    color=red,
    mark=triangle,
    ]
    coordinates {
    (40, 0.947)
    (60, 0.640)
    (80, 0.446)
    (100,0.216)
    (120, 0.149)
    (140,0.079)
    };
    
    \addplot+[smooth][
    color=orange,
    mark=*,
    ]
    coordinates {
    (60, 0.938)
    (80, 0.645)
    (100,0.474)
    (120, 0.239)
    (140, 0.138)
    (160, 0.062)
    };
    
    \addplot+[smooth][
    color=teal,
    mark=oplus,
    ]
    coordinates {
    (80,0.901)
    (100,0.727)
    (120,0.525)
    (140,0.278)
    (160,0.183)
    (180,0.087)
    };
    
    \addplot+[smooth][
    color=pink,
    mark=*,
    ]
    coordinates {
    (100,0.924)
    (120,0.759)
    (140,0.422)
    (160,0.334)
    (180,0.249)
    (200,0.154)
    (220,0.075)
    };
    
    \addplot+[smooth][
    color=purple,
    mark=square*
    ]
    coordinates{
    (120,0.951)
    (140,0.895)
    (160,0.704)
    (180,0.605)
    (200,0.462)
    (220,0.344)
    (240,0.243)
    (260,0.123)
    (280,0.098)
    (300,0.075)
};

    \legend{$f=0.01$, $f=0.08$, $f=0.15$, $f=0.30$, $f=0.50$, $f=0.80$}
\end{axis}
\end{tikzpicture}
\caption{Failure rate of chain quality for various $f$}
\label{cq_f}
\end{figure}

\begin{figure}[h!]
\centering
\begin{tikzpicture}
\begin{axis}[
    title={},
    xlabel={$k$},
    ylabel={Failure Rate},
    xmin=1.8, xmax=12.2,
    ymin=0, ymax=1,
    xtick={2,4,6,8,10,12},
    ytick={0,.20,.40,.60,.80,1},
    legend pos=outer north east,
    ymajorgrids=true,
    grid style=dashed,
    width=0.9\linewidth, height=6cm
]
\addplot+[smooth][
    color=blue,
    mark=square,
    ]
    coordinates {
    (2, 0.921)
    (3, 0.272)
    (4, 0.050)
    (5, 0.049)
    (6, 0.049)
    (7, 0.049)
    (8, 0.049)
    (9, 0.049)
    (10, 0.049)
    (11, 0.049)
    (12, 0.049)
    };
    \addplot+[smooth][
    color=red,
    mark=triangle,
    ]
    coordinates {
    (2, 0.951)
    (3, 0.938)
    (4, 0.3613375)
    (5, 0.084)
    (6, 0.049)
    (7, 0.049)
    (8, 0.049)
    (9, 0.049)
    (10, 0.049)
    (11, 0.049)
    (12, 0.049)
    };
    \addplot+[smooth][
    color=blue,
    mark=*,
    ]
    coordinates {
    (2, 0.951)
    (3, 0.951)
    (4, 0.712)
    (5, 0.206)
    (6, 0.049)
    (7, 0.049)
    (8, 0.049)
    (9, 0.049)
    (10, 0.049)
    (11, 0.049)
    (12, 0.049)
    };
    \addplot+[smooth][
    color=magenta,
    mark=x,
    ]
    coordinates {
    (2, 0.951)
    (3, 0.951)
    (4, 0.950)
    (5, 0.446)
    (6, 0.102)
    (8, 0.049)
    (9, 0.049)
    (10, 0.049)
    (11, 0.049)
    (12, 0.049)
    };
    \addplot+[smooth][
    color=teal,
    mark=oplus,
    ]
    coordinates {
    (2, 0.951)
    (3,0.951)
    (4, 0.951)
    (5, 0.792)
    (6, 0.300)
    (7, 0.067)
    (8, 0.049)
    (9, 0.049)
    (10, 0.049)
    (11, 0.049)
    (12, 0.049)
    };
    
    \addplot+[smooth][
    color=purple,
    mark=star,
    ]
    coordinates {
    (2, 0.951)
    (3, 0.951)
    (4, 0.951)
    (5, 0.950)
    (6,0.655)
    (7, 0.197)
    (8, 0.091)
    (9, 0.049)
    (10, 0.049)
    (11, 0.049)
    (12, 0.049)
    };
    
    \addplot+[smooth][
    color=teal,
    mark=|,
    ]
    coordinates {
    (2, 0.951)
    (3, 0.951)   
    (4, 0.951)
    (5, 0.951)
    (6, 0.938)
    (7, 0.537)
    (8, 0.170)
    (9, 0.050)
    (10, 0.049)
    (11, 0.049)
    (12, 0.049)
    };
    
    \addplot+[smooth][
    color=cyan,
    mark=o,
    ]
    coordinates{
    (2,0.951)
    (3,0.951)
    (4,0.951)
    (5,0.951)
    (6,0.951)
    (7,0.951)
    (8,0.796)
    (9,0.443)
    (10,0.152)
    (11,0.087)
    (12,0.049)
    };
    
    \legend{$f=0.01$, $f=0.08$, $f=0.15$,
    $f=0.23$,
    $f= 0.30$,
    $f= 0.39$,
    $f= 0.48$,
    $f= 0.60$}
\end{axis}
\end{tikzpicture}
\caption{Failure rate of common prefix for various $f$}
\label{cp_f}
\end{figure}
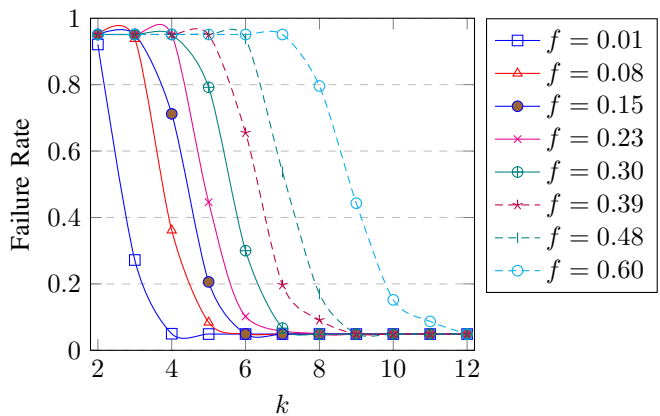

Changes in block propagation speed are not considered.  Since a round is a period of complete block propagation, $f$ is subject to change when network speeds change. In Bitcoin, block propagation speeds have rapidly increased over the last few years \cite{KIT2020}. For example, the time for a block to reach 99\% of nodes has decreased from 11 seconds to 2 seconds.\footnote{ Over the period from 2017 to 2020.} It follows that $f$, a key  parameter, has decreased as well. Because $f$ can change unexpectedly, measuring a cryptocurrency’s behavior over a range of $f$ values is important. 

\begin{figure}[t]
\begin{tikzpicture}
\begin{axis}[
xtick={2,3,4,5,6,7,8,9,10},
ytick={0.25, 0.50, 0.75},
yticklabel style={
        /pgf/number format/fixed,
        /pgf/number format/precision=2
},
scaled y ticks=false,
xlabel=$k$,
ylabel=$f$,
xmax = 10.5,
zlabel={Failure Rate},
width=0.9\linewidth, height=6cm
]
\addplot3[surf,mesh/rows=9] coordinates 
{
(2,0.013,0.92)	(2,0.081,0.951)	(2,0.152,0.951)	(2,0.226,0.951)	(2,0.304,0.951)	(2,0.386,0.951)	(2,0.475,0.951)	(2,0.6,0.951)	(2,0.7,0.951)	(2,0.8,0.951)
(3,0.013,0.272)	(3,0.081,0.887)	(3,0.152,0.951)	(3,0.226,0.951)	(3,0.304,0.951)	(3,0.386,0.951)	(3,0.475,0.951)	(3,0.6,0.951)	(3,0.7,0.951)	(3,0.8,0.951)
(4,0.013,0.049)	(4,0.081,0.311)	(4,0.152,0.711)	(4,0.226,0.95)	(4,0.304,0.951)	(4,0.386,0.951)	(4,0.475,0.951)	(4,0.6,0.951)	(4,0.7,0.951)	(4,0.8,0.951)
(5,0.013,0.049)	(5,0.081,0.033)	(5,0.152,0.205)	(5,0.226,0.446)	(5,0.304,0.792)	(5,0.386,0.95)	(5,0.475,0.951)	(5,0.6,0.951)	(5,0.7,0.951)	(5,0.8,0.951)
(6,0.013,0.048)	(6,0.081,0.048)	(6,0.152,0.048)	(6,0.226,0.101)	(6,0.304,0.299)	(6,0.386,0.655)	(6,0.475,0.937)	(6,0.6,0.951)	(6,0.7,0.951)	(6,0.8,0.951)
(7,0.013,0.048)	(7,0.081,0.048)	(7,0.152,0.048)	(7,0.226,0.048)	(7,0.304,0.066)	(7,0.386,0.197)	(7,0.475,0.537)	(7,0.6,0.951)	(7,0.7,0.951)	(7,0.8,0.951)
(8,0.013,0.048)	(8,0.081,0.048)	(8,0.152,0.048)	(8,0.226,0.048)	(8,0.304,0.048)	(8,0.386,0.091)	(8,0.475,0.169)	(8,0.6,0.796)	(8,0.7,0.951)	(8,0.8,0.951)
(9,0.013,0.048)	(9,0.081,0.048)	(9,0.152,0.048)	(9,0.226,0.048)	(9,0.304,0.048)	(9,0.386,0.048)	(9,0.475,0.049)	(9,0.6,0.443)	(9,0.7,0.916)	(9,0.8,0.951)
(10,0.013,0.048)	(10,0.081,0.048)	(10,0.152,0.048)	(10,0.226,0.048)	(10,0.304,0.048)	(10,0.386,0.048)	(10,0.475,0.048)	(10,0.6,0.152)	(10,0.7,0.721)	(10,0.8,0.951)
};
\end{axis}
\end{tikzpicture}
    \caption{Overview of parameter space for common prefix property}
    \label{3d_f}
\end{figure}
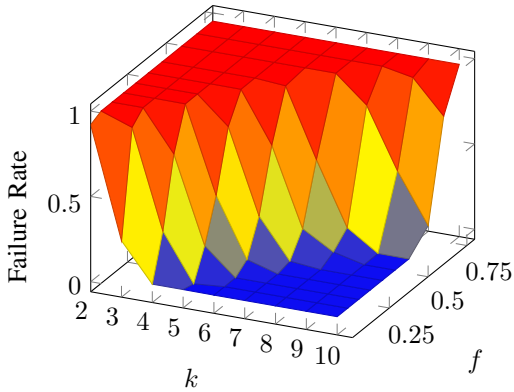

For the following experiments, we let $n = 8, \alpha = 0.33 , \text{and } \mu =0.39$. 
Over a range of $f$ values, we find that Bitcoin has very different security properties. For smaller values of $f$, the failure rate of the chain quality and common prefix properties converge faster (Figures \ref{cq_f}, \ref{cp_f}, \ref{3d_f}). When the failure rate of chain quality converges faster, then we can look at a shorter length of blocks and be confident there are $\mu\ell$ honest blocks. When the common prefix converges faster, we can wait for less block confirmations and be confident that double spending will not occur. While these behaviors help with security, they do not help with transaction processing speed. To improve transaction processing speed, a cryptocurrency must increase its block generation rate \cite{eprint-2015-26827}.

Decreasing block generation times provide an opportunity for increased transaction throughput. If a cryptocurrency can improve its block propagation speed enough, then the block generation rate can be decreased such that $f$ stays constant. The behavior of the chain quality and common prefix properties will be the same, but blocks and transactions will be produced more rapidly. 

In the case of Bitcoin, this raises a question: should cryptocurrencies aim for idealogical or security based consistency? While Bitcoin block generation rate is kept at 10 minutes, this does not guarantee consistency from the backbone perspective. Bitcoin, from the perspective of the backbone protocol, has changed greatly over the last few years.

\section{Related Work}
\cite{Chaudhary_2015} and \cite{10.1007/978-3-319-77935-5_11} use UPPAAL SMC to model Bitcoin, but do not build off the backbone protocol. \cite{Chaudhary_2015} analyzes double spending, whereas \cite{10.1007/978-3-319-77935-5_11} analyzes an Andresen attack. Neither model chain quality or chain growth. 

\cite{gervais2016security} provide a framework for modeling the security of PoW blockchains using a Markov Decision Process. They also show that selfish mining is not  always a rational strategy. Still, their rational attacker model does not account for incentives outside of blockchain. For example, an attacker may lose bitcoin during an attack, but have a payoff in USD. \footnote{Through financial derivatives or increased market share.}

In \cite{Sapirshtein2015OptimalSM} the authors measure the effects of uniform tie breaking on the profit threshold for selfish mining.  By looking at the revenue of the adversary, \cite{Sapirshtein2015OptimalSM} implicitly measures the chain quality property from \cite{garay_bitcoin_2015}.  Their results showed that uniform tie breaking limited the power of strongly communicating attackers, but enhanced the power of poorly communicating attackers. Still, no work was done to directly measure the effects of this modification on the common prefix and chain growth properties.

\section{Future Work}
Some avenues for future work are:

\begin{itemize}[leftmargin=0.25cm]
    \item \textbf{Other adversarial strategies}: Our model can be used to model other adversarial strategies, besides selfish mining.
    \item \textbf{Delay-bounded model}: Our model can be extended to the delay-bounded version of the backbone protocol, where rounds are not highly synchronous, but there is a bound on the time it takes for messages to be delivered.
    \item \textbf{Honest Mining Pools}: Our work assumes all honest parties have the same hashing capabilities. In reality, different parties and pools have different hashing capabilities. Future work could account for this. This would capture the importance of certain parties in terms of block propagation. For example, it would be more beneficial for a selfish miner to quickly relay his block to a large mining pool than to a single miner with a negligible percentage of the total hash rate
\end{itemize}

\section{Conclusion}
This paper presents a case-study of model checking PoW cryptocurrencies using the Bitcoin Backbone Protocol as a foundation. 
We show how to model the protocol using Statistical Model Checking tools, and identify concrete security properties of the protocol. 
We use the model to demonstrate how design decisions can impact different concrete backbone protocol properties in different ways, 
in a manner that is not obvious from prior asymptotic analysis. 

This paper attempts to explain the value of applying the foundation introduced by \cite{garay_bitcoin_2015} to practice. 
The experiments above map out the effectiveness of a selfish mining strategy against various deployment parameters.  
By doing this, we are able to derive results that lay a direction for further work in the design and analysis of PoW protocols.

 \section{Acknowledgments}
The authors thank Marco Patrignani (Stanford University) and Avradip Mandal, Hart Montgomery, Arnab Roy (Fujitsu Laboratories of America Inc.) for their assistance and insight in formulating the ideas underlying our model and results. The authors thank the Office of Naval Research for support through grant N00014-18-1-2620, Accountable Protocol Cus tomization.
\printbibliography

\clearpage
\newpage
\section{Modeling the Bitcoin Backbone Protocol in UPPAAL}

In this section, we provide details on how we modeled the Bitcoin Backbone Protocol with UPPAAL-SMC.

\subsection{Global Declarations}

To model the backbone protocol, we designed three custom data structures: Block, Global\_Ledger, and Diffusion.

\begin{declaration}
  typedef struct \{ \\
    $\;$ int[0, node\_max] id; \\
    $\;$ int[0, total\_run\_time] rd; \\
    $\;$ int parent; \\
    $\;$ int block\_num; \\
    $\;$ int length; \\
    $\;$ bool sent\_to[node\_max]; \\
    $\;$ int[0, 1] is\_private; \\
    $\;$ int num\_adv\_blocks;   \\
    \} Block;
  \caption{Block}
  \label{block}
\end{declaration}

\begin{declaration}
  typedef struct \{  \\
    $\;$ Block blockchain[block\_max];    \\
    $\;$ int best\_block[node\_max];   \\
    $\;$ int max\_len;    \\
    \} Global\_Ledger;
  \caption{Global\_Ledger}
  \label{ledger}
\end{declaration}

\begin{declaration}
  typedef struct \{  \\
    $\;$ int receive[node\_max][node\_max];   \\
    $\;$ int receive\_len[node\_max];  \\
    $\;$ int to\_be\_diffused[node\_max];   \\ 
    $\;$ int[0, node\_max] to\_be\_diffused\_len;   \\
    \} Diffusion;
  \caption{Diffusion}
  \label{diffusion}
\end{declaration}

Declaration \ref{block} shows the structure of a block in our model.  Each block has an unique identifier, a round number (indicating the round it was created in), a parent id that indicates the previous block in the chain, and a block number indicating the number of blocks ever created at the time of its creation. Each block also stores information such as its depth from the genesis block, an array of participants that the block was sent to, a flag indicating whether it is held private, and the number of blocks created by adversaries in its chain.

Declaration \ref{ledger} shows the structure of the global blockchain ledger. Global\_ledger keeps track of every block created in the network with blockchain[].  Miners maintain their local chain by pointing to a block in best\_block[]. Finally, max\_len tracks the length of the longest chain(s) in the network.

Declaration \ref{diffusion} shows the structure of the diffusion model. Diffusion models the diffusion functionality used in the backbone protocol. A 2D array, mimicking the RECEIVE() tape from the backbone, is used to keep track of block propagation.

\subsection{Honest and Adversarial Parties}

\begin{figure}[h!]
    \centering
    \includegraphics[width=0.5\textwidth]{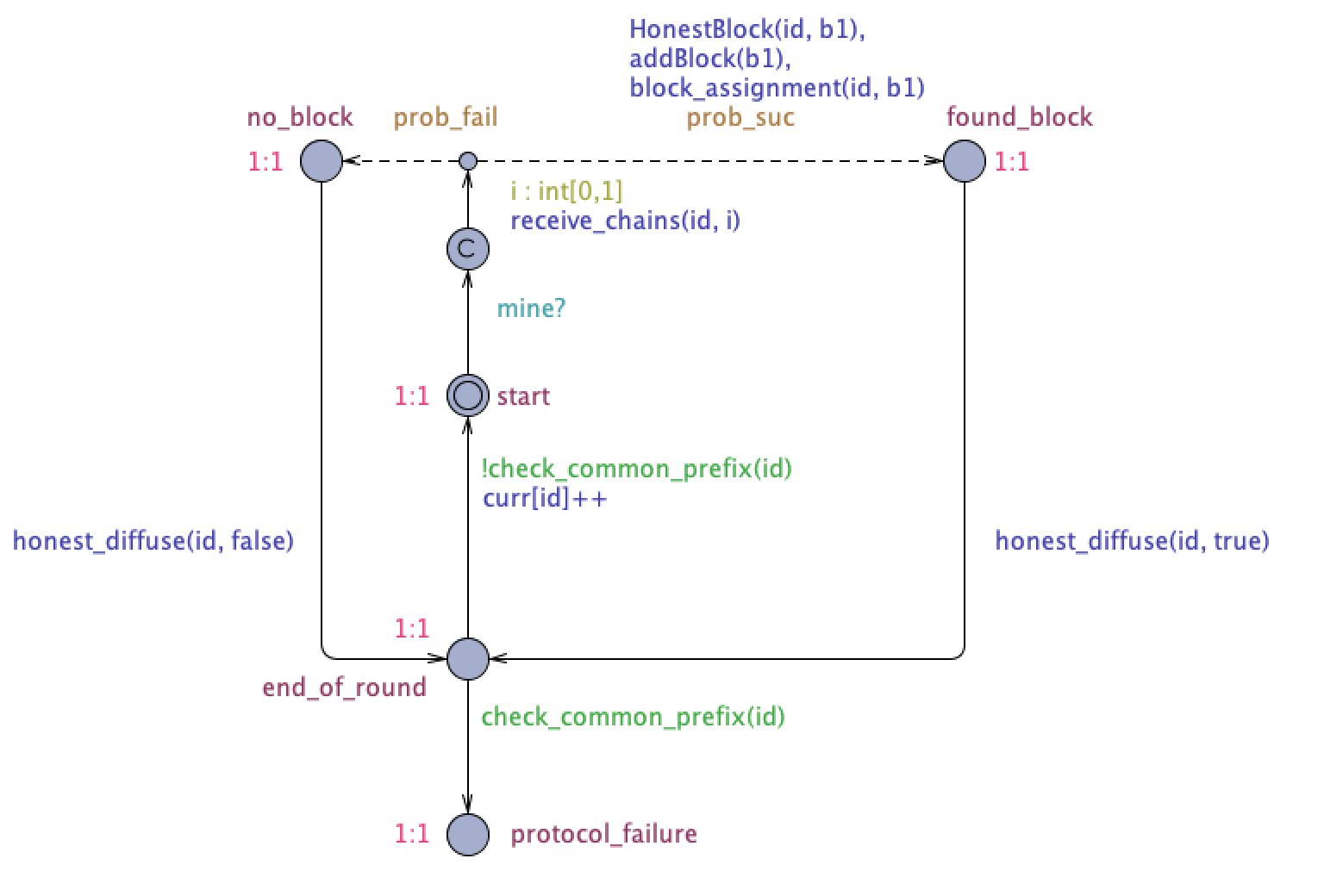}
    \caption{An honest party's state diagram}
    \label{fig:honest}
\end{figure}

\begin{figure}[h!]
    \centering
    \includegraphics[width=0.5\textwidth]{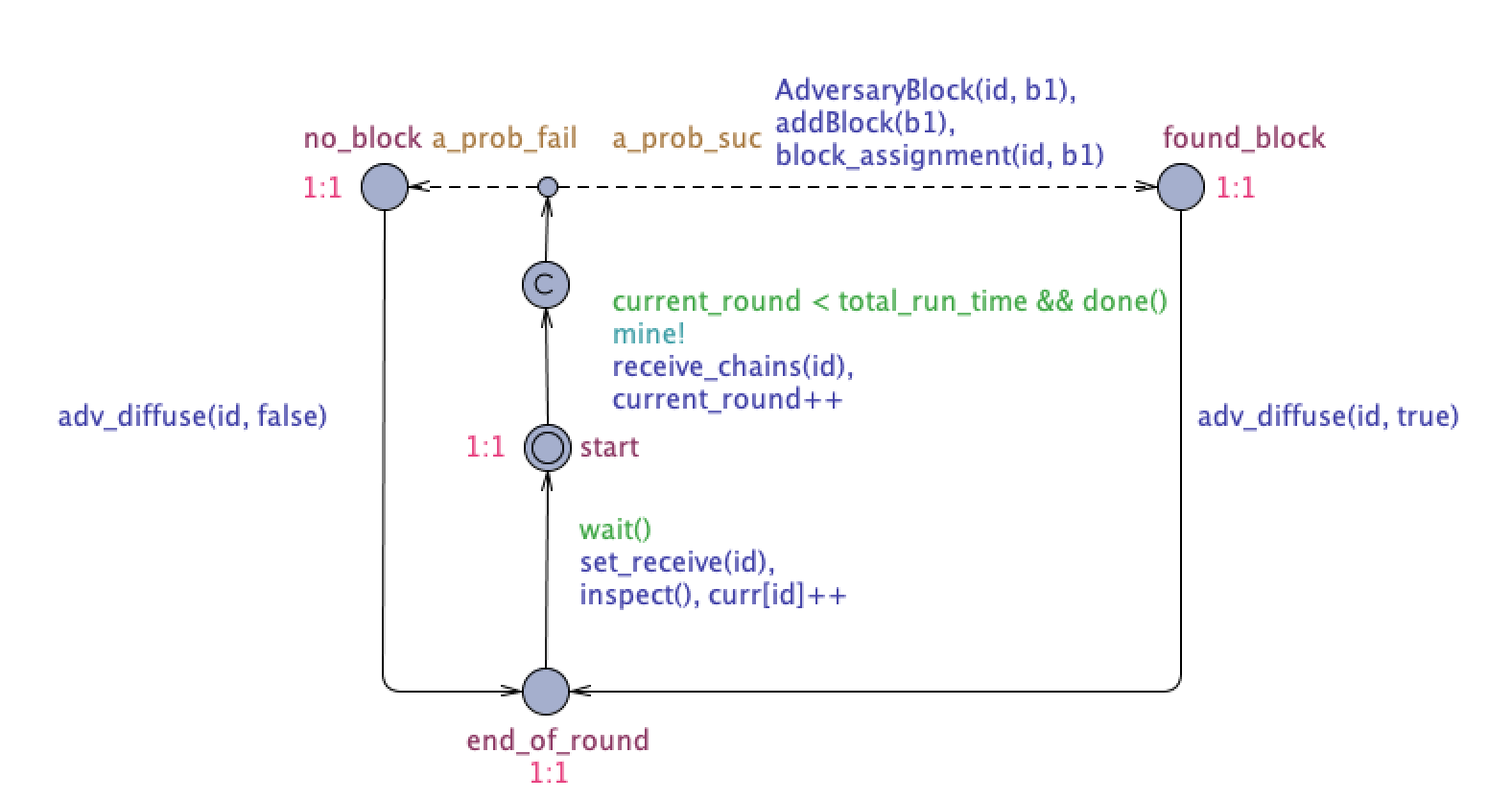}
    \caption{The adversary's state diagram}
    \label{fig:adv}
\end{figure}

As shown in in Figure \ref{fig:honest} and Figure \ref{fig:adv}, we modeled honest parties and the adversary  separately. Each state diagram consists of five non-trivial states:

\begin{enumerate}
    \item \textbf{Start}: the start state of each round
    \item \textbf{End}: the end state of each round
    \item \textbf{Protocol Failure}: indicates a failure of one of the backbone properties
    \item \textbf{No Block}: indicates a party's mining outcome is unsuccessful
    \item \textbf{Found Block}: indicates a party's mining outcome is successful
\end{enumerate}

Each round begins at the \textbf{start} state and ends at the \textbf{end\_of\_round} state. The \textbf{protocol\_failure} state represents a failure in the backbone protocol. The \textbf{no\_block} and \textbf{found\_block} states correspond to a miners PoW outcome in the current round. 

We note features of the state diagrams:
\begin{itemize}
    \item One or more backbone property is verified at the end of each round. In this example, the expression \textbf{check\_common\_prefix} will force a miner to enter the failure state when the common prefix is broken.
    \item The probability that at least one honest party succeeds in finding a PoW solution in a round ($f$), is captured with probabilistic edges. A weight assigned to each edge is used to vary $f$. 
    \item The synchronization channels \textbf{mine!} and \textbf{mine?} prevent miners from starting a new round before everyone has finished the previous round.
\end{itemize}

\subsection{Property Checking Algorithms}

At the end of each round, we check each honest chain against one or more of the backbone properties. If a party's chain does not satisfy the backbone property, the party enters the failure state of the model. UPPAAL will terminate and consider this run a failure. 

We illustrate how property is checked:
\begin{itemize}

\item \textbf{Common Prefix}: 
Common prefix checks that the block $k$ deep in an honest chain is in every other honest chain. This search is pruned by ignoring parties that point to the same \textbf{best\_block}. This means they have identical chains.

\item \textbf{Chain Quality}:
Chain quality is checked by counting the blocks in honest chains that were contributed by honest parties. The share of these contributions over any set of $\ell$ blocks should be at least $\mu$$\ell$.

\item \textbf{Chain Growth}:
Chain growth is checked by iterating over the blocks of an honest chain until a block $s$ rounds or older is found. Chain growth is satisfied if at least $\tau \cdot s$ blocks were found in this time frame.   
\end{itemize}

\subsection{Selfish Mining}
As shown in Table 6, each block has a data field \textbf{is\_private}. Our adversary deterministically chooses whether to keep their blocks private. There are two cases:
\begin{enumerate}
  \item If the adversary's private chain is at least one block ahead of the longest honest chain, it will release one block from its private chain for every honest block that is published. 
  \item When the adversary's private branch is depleted, it will return to mining on the public branch.
  \end{enumerate}

\end{document}

%% file: mymacro.tex
\newenvironment{declaration}[1][htb]
  {
   \begin{algorithm}[#1]%
  }{\end{algorithm}}